\documentclass[fleqn,usenatbib]{mnras}

\usepackage[T1]{fontenc}
\usepackage{graphicx}

\DeclareRobustCommand{\VAN}[3]{#2}
\let\VANthebibliography\thebibliography
\def\thebibliography{\DeclareRobustCommand{\VAN}[3]{##3}\VANthebibliography}

\usepackage{graphicx}	
\usepackage{amsmath}	
\usepackage{amssymb}	
\usepackage{multirow}

\title[GCs in S-PLUS]{A new method to detect globular clusters with the S-PLUS survey}

\author[M. L. Buzzo et al.]{Maria Luísa Buzzo,$^{1,2}$\thanks{E-mail: luisa.buzzo@gmail.com}
Arianna Cortesi,$^{3,10}$
Duncan A. Forbes,$^{1}$
Jean P. Brodie,$^{1,4}$
Warrick J. Couch,$^{1}$
\newauthor
Carlos Eduardo Barbosa,$^{2}$
Danielle de Brito Silva,$^{5}$
Paula Coelho,$^{2}$
Ana L. Chies-Santos, $^{6,7}$
\newauthor
Carlos Escudero,$^{8,9}$
Leandro Sesto,$^{8,9}$
Kar\'in {Men\'endez-Delmestre},$^{3}$
Thiago S. Gon\c{c}alves,$^{3}$
\newauthor
Cl\'ecio R. Bom,$^{10,11}$
Alvaro {Alvarez-Candal},$^{12,13,14}$
Anal\'ia V. Smith Castelli,$^{8,9}$
William Schoennell,$^{15}$
\newauthor
Antonio Kanaan,$^{16}$
Tiago Ribeiro,$^{17}$
Claudia Mendes de Oliveira$^{2}$\\
$^{1}$ Centre for Astrophysics \& Supercomputing, Swinburne University, Hawthorn VIC 3122, Australia \\
$^{2}$ Universidade de S\~ao Paulo, IAG, Rua do Mat\~ao 1226, Cidade Universit\'aria, S\~ao Paulo 05508-900,Brazil\\
$^{3}$ Observat\'orio do Valongo, Ladeira do Pedro Ant\^onio 43, CEP:20080-090, Rio de Janeiro, RJ, Brazil\\
$^{4}$ University  of California Observatories, Santa Cruz, CA 95064, USA\\
$^{5}$ N\'ucleo de Astronom\'ia, Universidad Diego Portales, Av. Ej\'ercito Libertador 441, Santiago, Regi\'on Metropolitana, Chile \\
$^{6}$ Instituto de F\'isica, Universidade Federal do Rio Grande do Sul (UFRGS), Av. Bento Gon\c{c}alves, 9500, Porto Alegre, RS, Brazil \\
$^{7}$ Shanghai Astronomical Observatory, Chinese Academy of Sciences, 80 Nandan Road, Shanghai 200030, China \\
$^{8}$ Facultad de Ciencias Astronómicas y Geofísicas, Universidad Nacional de La Plata, Paseo del Bosque s/n, B1900FWA, La Plata, Argentina \\
$^{9}$ Instituto de Astrof\'isica de La Plata (CCT La Plata - CONICET - UNLP), B1900FWA, La Plata, Argentina \\
$^{10}$ Centro Brasileiro de Pesquisas F\'isicas, Rua Dr. Xavier Sigaud 150, CEP 22290-180, Rio de Janeiro, RJ, Brazil\\
$^{11}$ Centro Federal de Educa\c{c}\~ao Tecnol\'ogica Celso Suckow da Fonseca, Rodovia M\'ario Covas, quadra J, CEP 23810-000,  Itagua\'i, RJ, Brazil\\
$^{12}$ Instituto de Astrof\'isica de Andaluc\'ia, CSIC, Apt 3004, E18080 Granada, Spain\\
$^{13}$ IUFACyT, Universidad de Alicante, San Vicent del Raspeig, E03080, Alicante, Spain\\
$^{14}$ Observat\'orio Nacional / MCTIC, Rua General Jos\'e Cristino 77, Rio de Janeiro, RJ, 20921-400, Brazil\\
$^{15}$ GMTO Corporation 465 N. Halstead Street, Suite 250 Pasadena, CA 91107, USA \\
$^{16}$
Departamento de F\'isica, Universidade Federal de Santa Catarina, Florian\'opolis, SC, 88040-900, Brazil \\
$^{17}$
NOAO, P.O. Box 26732, Tucson, AZ 85726, USA \\}

\date{Accepted 2021 November 26. Received 2021 November 25; in original form 2021 August 24}

\pubyear{2021}

\begin{document}
\label{firstpage}
\pagerange{\pageref{firstpage}--\pageref{lastpage}}
\maketitle

\begin{abstract}
In this paper, we describe a new method to select globular cluster (GC) candidates, including galaxy subtraction with unsharp masking, template fitting techniques and the inclusion of Gaia's proper motions. We report the use of the 12-band photometric system of S-PLUS to determine radial velocities and stellar populations of GCs around nearby galaxies. Specifically, we assess the effectiveness of identifying GCs around nearby and massive galaxies (D $< 20$ Mpc and $\sigma > 200$ km/s) in a multi-band survey such as S-PLUS by using spectroscopically confirmed GCs and literature GC candidate lists around the bright central galaxy in the Fornax cluster, NGC 1399 (D = 19 Mpc), and the isolated lenticular galaxy NGC 3115 (D = 9.4 Mpc). Despite the shallow survey depth, that limits the present work to $r < 21.3$ mag, we measure reliable photometry and perform robust SED fitting for a sample of 115 GCs around NGC 1399 and 42 GCs around NGC 3115, recovering radial velocities, ages, and metallicities for the GC populations.
\end{abstract}




\section{Introduction}
\label{sec:introduction}
Large-area imaging surveys have become increasingly important in the past years to allow simultaneous characterisation of large samples of objects such as globular cluster (GC) candidates, both bound and unbound to galaxies \citep{Lee10, Peng11}. The advantages are many, e.g., such surveys deliver homogeneous datasets over large continuous areas of the sky - they may cover entire nearby clusters of galaxies \citep[e.g., Fornax Deep Survey, ][]{Cantiello20} - allowing the study of the outskirts of the galaxies, not usually possible in pointed observations. From such data, comprehensive catalogues of GC candidates have been extracted around galaxies inhabiting a wide range of environments, redshifts and masses \citep{Ko19}. 

GC candidates can usually be separated from other contaminant sources, such as stars, compact galaxies and high-redshift galaxies, using a combination of photometric (e.g., expected magnitudes and colours) and morphometric (e.g., concentration index, roundness, elongation) properties. Yet, spectroscopic follow-up is required to confirm the real nature of GC candidates by measuring their radial velocities \citep{Brodie}. Alternatively, images taken with space-based telescopes allow these objects to be spatially resolved, to measure their sizes \citep{Larsen01} and to separate them from point-like sources.

Nonetheless, both techniques to confirm the nature of GCs, i.e., obtaining radial velocities through spectroscopy or measuring sizes through space-based imaging, are limited, given that they are possible for a relatively small number of targets and they are restricted to the GCs close to the parent galaxy. To this is added the typically small fields of view (FoV) of available instruments or the difficulty in choosing suitable GC candidates for spectroscopy in the outskirts of the galaxies, where the ratio of GCs over stars is small. It is worth mentioning, nonetheless, that future missions such as Euclid \citep{Bates19,Lancon21}, the Nancy Grace Roman Space Telescope \citep{Spergel15,Troxel20} and the Chinese Space Station Telescope \citep[CSST, ][]{Zhang19,Zhou21} will provide imaging and spectroscopy for unprecedented large fields-of-view in different wavelength regimes, allowing the study and confirmation of GC populations out to very large radii.

Ground-based imaging surveys with poor spatial resolution and shallow depths are not capable of confirming the nature of GCs. Still, extensive coverage of the sky allows the identification of GC candidates in different environments and distances. This renders possible, on the one hand, statistical estimates of the number and specific frequencies of GC systems \citep{Prole_19} and, on the other hand, to provide reliable lists of candidates for follow-up spectroscopic studies \citep{Forbes17}.

In this work, we describe a new method to search, catalogue and characterise GC candidates using data from the Southern Photometric Local Universe Survey \citep[S-PLUS, ][]{MendesdeOliveira19} internal Data Release 3 (iDR3). We show that the spectral energy distributions (SED) of nearby GCs around bright galaxies can be well determined along with radial velocities and stellar populations of GCs using the 12-band optical images from S-PLUS (five in Sloan-like filters and seven narrow-band filters).

As test cases, we use two early-type galaxies with both spectroscopically confirmed catalogues of GCs and literature lists of GC candidates, NGC 1399 (D = 19 Mpc) and NGC 3115 (D = 9.4 Mpc), to evaluate the limitations of the survey and the feasibility of obtaining GCs stellar population properties by making use of the unique 12-band filter system used in S-PLUS (and its northern counterpart J-PLUS, \citealt{Cenarro19}). In fact, a similar study of GC candidates populations using J-PLUS is pursued by Brito-Silva et al. submitted, targeting the S0 galaxy NGC 1023.

The paper is structured as follows: in \textsection \ref{sec:data}, we describe the data used in this study; in \textsection \ref{sec:GC}, we detail our GC selection method; in \textsection \ref{sec:params}, we describe the measured parameters; and in \textsection \ref{sec:testcases}, we apply our method to the two test-galaxies and discuss our results.

\section{Data}
\label{sec:data}
S-PLUS is a 12-filter imaging survey that uses a robotic 0.8-meter telescope located at the Cerro Tololo Inter-American Observatory (CTIO) in Chile. Aiming at covering $\sim9300$ square degrees of the Southern sky, S-PLUS has a pixel scale of $0.55$ pixel per arcsec and a FoV of 1.96 deg$^2$. It uses the Javalambre filter system \citep{Java}, including 5 broad bands ($u$, $g$, $r$, $i$, $z$), and 7 narrow bands centred at important spectral features of astronomical sources (e.g., [OII], Ca H+K, D4000, H$\delta$, Mg$b$, H$\alpha$ and CaT). The S-PLUS survey reaches an average 3$\sigma$ depth of $r\sim 21.3$ mag, and although shallow, its power relies on the inclusion of the seven narrow-band filters that allow for precise SED fitting results. 
In this work, we use data from the S-PLUS internal Data Release 3 (iDR3), where the data reduction and data calibration follow the pipeline developed for S-PLUS DR2, thoroughly described in \cite{Almeida-Fernandes21} (hereafter AF21).

\subsection{Initial photometric measurements}

As described in AF21, S-PLUS has an entire pipeline focused on the detection and creation of photometric catalogues. This pipeline, nonetheless, was not designed to find faint and compact objects such as extragalactic globular clusters. Therefore, with very few exceptions, looking for extragalactic GCs in the S-PLUS catalogues is not the ideal way of finding these objects within the survey. In this work, we develop a new pipeline to find and extract the photometry of these sources in the S-PLUS images using {\sc DAOfind} and aperture photometry.
Below we carefully explain the identification process and extraction of the photometry of the sources. 

\subsubsection{Galaxy subtraction}

We compare three methods to perform galaxy-light subtraction, an important step to identify and reliably obtain photometry of GCs that lie within the central region of galaxies. These methods are: 1) galaxy fitting with GALFITM \citep{Haussler2013} using one component, 2) galaxy fitting using the package {\sc Isophote} from the {\sc Photutils} \citep{photutils} library in Python, and 3) unsharp masking, an image sharpening technique to smooth the original image, enhancing the contrast of fine structure \citep{Malin}. The unsharp masks used in this work were created using a 25-by-25 pixels median box and a circular Gaussian smoothing with $\sigma = 5$ pixels.

\begin{figure*}
    \centering
    \includegraphics[width=\textwidth]{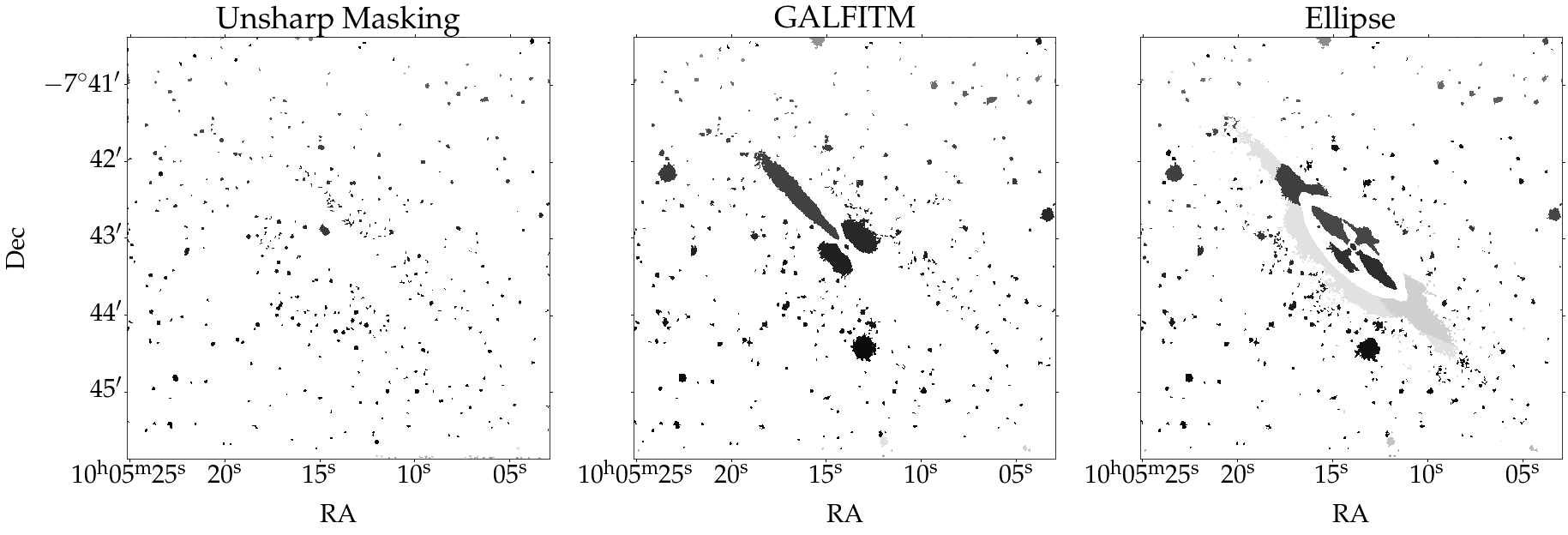}
    \caption{Comparison of residual images after applying methods of galaxy subtraction in NGC 3115. Left: Unsharp masking using a 25x25 px median box and $\sigma$=5 px. Middle: Fitting with GALFITM, using a single component (Single S\'ersic model). Right: Fitting with {\sc Isophote} code from the photutils library to exclude the galaxy light and provide final images to find the globular cluster candidates.}
    \label{fig:unsharp}
\end{figure*}

Amongst our tests, the unsharp masking method delivered much cleaner final images. This is because unsharp masking is an image smoothing technique and is different from methods of galaxy fitting, such as {\sc GALFITM} and {\sc Isophote.} These galaxy fitting methods are strongly dependent on the number of components fitted and how well the models describe the galaxy light profile, making it hard to apply them to complex systems or to large samples.
The comparison of the three  galaxy-subtraction methods is shown in Fig. \ref{fig:unsharp}, where we can clearly see that, while unsharp masking is capable of fully excluding the main light of the galaxy, the methods of galaxy fitting, such as GALFITM and {\sc Isophote}, leave behind residuals of other unfitted galaxy components.
From a quantitative point-of-view, for our test-case galaxy NGC 3115 \citep{Pota13}, unsharp masking increased by 22\% the sample of spectroscopically confirmed GCs identified in S-PLUS, while GALFITM increased it by 14\% and {\sc Isophote} by 12\% when compared to the sample of GCs found without applying any method of galaxy subtraction. 
This indicates that galaxy subtraction substantially expands the number of objects found and it is, therefore, the best method for this task. 

\subsubsection{{\sc DAOfind} and aperture photometry}
\label{sec:identification}
For the source detection, we use {\sc DAOfind} from the {\sc Photutils} library. For this, we defined a threshold of 2$\sigma$ above the background to detect sources and a FWHM of $\sim 3$ pixels ($\sim 1.6$ arcsec). The detections were performed in a detection image, generated by combining the $g$, $r$, $i$ and $z$ images from S-PLUS.

Once we find the sources that meet these criteria, we run the {\sc aperture\_photometry} code from {\sc Photutils} in each individual image. Since S-PLUS has a pixel size of 0.55 arcsec and seeing of $\sim 1.5$ arcsec, we assume point sources should have most of their light within $\sim 3$ arcsec. Therefore, for the aperture photometry, we use a slightly smaller aperture of 2 arcsec diameter to avoid the contamination of nearby sources in the GC candidate photometry, as a compromise to optimize the SNR and get rid of the galaxy background light (see also AF21 for a further discussion of S-PLUS photometry). Then we perform local sky subtraction, placing an annulus of 1-arcsec width around each source, with an inner radius of 3 arcsec and outer radius of 4 arcsec, followed by a sigma clipping until convergence to the best background value was reached. 

Later, we add in an aperture correction calculated using the growth curves of isolated GC candidates.
To do that, we calculate the average magnitude of five GCs located in the outermost regions of the galaxy, where the diffuse light of the latter is sufficiently faint so as to not compromise the determination of the GC candidates’ flux. We use this to create the growth curves out to an aperture of 30 arcsec, and identify the radius at which such magnitude reaches a plateau; this turns out to be approximately 6 arcsecs. Then, we calculate the average difference ($\Delta$m) between the magnitude embedded within a radius of 2 arcsec and the magnitude at 6 arcsec. The aperture correction ($\Delta$m) is then added to the magnitude extracted at 2 arcsec for every GC candidate.

As the last step, we corrected the data for Galactic extinction using the correction derived by \cite{Cardelli}. 
In this manner, we define all identified point sources as our parent sample of GC candidates.


\section{Globular cluster selection}
\label{sec:GC}

In this work, we employ a new method to select GC candidates. We use a combination of known selection criteria (i.e., expected magnitudes and concentration) with newly proposed selection methods, including template fitting techniques and the inclusion of Gaia's proper motions, to generate our GC candidate lists. 
Starting with all point source detections (identified following Section \ref{sec:identification}) within a FoV 50 times bigger than the effective radius of the studied galaxies, we analyze the impact of each of the following cuts on the total number of detections.
Below we describe each of the selection criteria.

\begin{itemize}
\item Magnitude (Mag$_{\rm bright}$): The cuts based on expected magnitudes follow the calculation of the 3$\sigma$ depth of the S-PLUS survey estimated to be on average $r \sim 21.3$ mag \citep{Almeida-Fernandes21}.

\item Concentration index \citep[$C$, ][]{Peng11}: We use the magnitude difference at the 3 arcsec aperture and 6 arcsec aperture to select the point-like sources, since for these objects, after applying the aperture corrections, $C$ is expected to be consistent with zero, assuming a scatter of 0.2 mag, to comply with the errors in the photometry of the sources.

\item Gaia proper motion (pm): One further step to exclude stars in our samples was to use Gaia’s EDR3 \citep[][complete down to $g\sim20$ mag]{GAIA_EDR3}. The principle is to use the proper motion ($\mu$) of the objects to separate definite stars from other objects. To do so, we use the signal-to-noise ratio of the proper motion \citep[SNR$_{\mu}$, ][]{Voggel20}:
\begin{equation}
  {\rm SNR}_{\mu} = \sqrt{\mu_{\rm RA}^2 + \mu_{\rm DEC}^2} \Biggm/ \sqrt{\sigma\mu_{\rm RA}^2 + \sigma\mu_{\rm DEC}^2}
\end{equation}

In this case, the non-stars are expected to have proper motions that are consistent with 0 at the $3\sigma$ confidence level, while genuine stars are expected to have SNR$_{\mu}>3$. This is an important step using S-PLUS, since due to its small point spread function, GCs usually get mistaken with stars rather than extended sources, such as ultra-compact dwarfs. Therefore, with this cut, we intend to exclude definite stars that may be contaminating the GC candidate list.

\item Template fitting (TF$_{\rm crit}$): Extragalactic GCs have statistically higher radial velocities than MW stars \citep{Forbes17}, which can be exploited to get yet another discrimination between sources.
We use {\sc LePhare} \citep{arnouts2}, a template fitting (TF) routine based on a $\chi^2$ minimization, to fit our data with both galactic and stellar templates. For this analysis, when the data are fitted by galactic templates, they will necessarily be accompanied by a radial velocity estimate. In contrast, in the fit with the stellar template, no radial velocity (RV) is returned. 
The primary templates used for this analysis are a set of galactic SEDs derived by the COSMOS survey collaboration \citep{cosmos}, and the Pickles stellar spectra library \citep{pickles}.
We select the objects that are better fitted by galactic templates than by stellar templates, allowing a 50\% margin of error ($\chi^2_{\rm red} {\rm (galactic)} < 1.5 \chi^2_{\rm red} {\rm (stellar)}$).
The GC candidates that meet this criterion can be understood as populating an area of the parameter space where objects are better fitted when we allow for a non-null radial velocity estimate rather than fixing RV to zero. 

\end{itemize}

\begin{table}
    \centering
    \caption{Morphometric and photometric GC selection criteria.}
    \begin{tabular}{lc} \hline
        \textbf{Method} & \textbf{Value} \\ \hline
        Mag$_{\rm bright}$ & $<3\sigma$ depth ($21.3$ mag) \\
        $C$ & $-0.2$ $<\, C \,<$ $0.2$ \\
        Gaia pm & SNR$_{\mu}$ $< 3$\\
        TF$_{\rm crit}$ & $\chi^2_{\rm red} {(\rm galactic)} < 1.5\chi^2_{\rm red} {(\rm stellar)}$\\ \hline
    \end{tabular}
    \label{tab:selectGCs}
\end{table}

We note that the combination of all of the methods to select GCs, summarised in Table \ref{tab:selectGCs}, provides the best results in all of our tests, as shown in Section \ref{sec:testcases}.

\section{Measured parameters}
\label{sec:params}

\subsection{Radial velocities}

The template fitting method not only helps to select GC candidates around the target galaxies but also allows for estimates of radial velocities (or photometric redshifts). Up until now, as mentioned in Section \ref{sec:introduction}, estimates of radial velocity were mainly derived from spectroscopy and are one of the ways to confirm GC populations. In this work, we recover radial velocities for all GC candidates that meet the criteria summarised in Table \ref{tab:selectGCs}. For the template fitting technique, we assume a standard $\Lambda$CDM model, with H$_0 = 70.5$ km/s/Mpc \citep{Komatsu}.

The RVs are recovered by computing synthetic photometric magnitudes out of these empirical spectra at various redshifts, and recording the redshift that best reproduces the observed photometry. The method to perform the template fitting and recover radial velocities for the entire S-PLUS iDR3 is described in Buzzo et al. in preparation.

\subsection{Stellar Populations}

For the objects that comply with our 4 combined selection criteria, we can perform robust SED fitting due to the constraints enabled by the extensive S-PLUS filter system, thus leading to reliable ensemble stellar population properties for the GCs \citep{Izaskun}. 
Typically, all 12 S-PLUS bands are used to perform the fits. Nonetheless, GCs with up to 5 missing bands were also fitted since we consider 7 bands to be still representative of the overall shape of the SED and provide good-quality fits.
%

We use the fully Bayesian Monte Carlo Markov Chain (MCMC)-based code {\sc prospector} \citep{Leja17}, coupled with the Flexible Stellar Population Synthesis package \citep[FSPS; ][]{Conroy09}, which allows fitting several parameters affecting the SED. 
For our analysis, we fitted three free parameters: stellar mass (M$_{\star}$), metallicity ([Fe/H]) and age since the first onset of star formation (t$_{\rm age}$). We fix the e-folding timescale ($\tau$) to 1 Gyr, assuming a star formation history with an early and single burst for the GCs \citep{Brodie}.

We placed priors on these three free parameters. The stellar mass could vary with M$_\star$ = 10$^{4-7}$ M$_{\odot}$, as per the typical masses of GCs \citep{Brodie}, while the metallicity and age we allow to vary within the entire allowed range in Prospector, since different GC populations could have different ages and metallicities \citep{Usher}, i.e., [Fe/H] = -2.0 to 0.2 dex, and t$_{\rm age} = 0.1-14$ Gyr.

We fit the detected GCs using an exponentially declining star formation history, a Kroupa initial mass function \citep[IMF, ][]{Kroupa} and a Calzetti \citep{Calzetti} extinction law. We fix the dust to zero and the redshift of the GCs to the recovered radial velocity from our template fitting technique. 

\section{Test-cases}
\label{sec:testcases}

To test the method, we use two early-type galaxies with confirmed catalogues of GCs, NGC 1399 and NGC 3115. For this test, we want to assess the success rate of both spectroscopically confirmed GCs and literature GC candidates found in S-PLUS after applying our selection cuts (Table \ref{tab:selectGCs}) and how well we can recover their physical parameters. 

\subsection{NGC 1399}
\label{sec:ngc1399}
 
\begin{table*}
     \centering
    \caption{GC candidate selection criteria applied to our two test case galaxies. The criteria are applied to the whole sample of detections within the FoV of the galaxies and compared with both the entire spectroscopic and photometric literature samples, as well as with these samples constrained to $r<21.3$ mag (in boldface).}
     \begin{tabular}{l|c|c|c|c|c|c} \hline
          & \multicolumn{3}{c}{NGC 1399} & \multicolumn{3}{c}{NGC 3115} \\ 
          & Total & GCs (spec) & GC candidates (phot) & Total & GCs (spec) & GC candidates (phot) \\ \hline
          Initial sample & -- & 575 & 1589 & -- & 122 & 781 \\
          \textbf{Initial sample (r $<21.3$ mag)} & \textbf{-} & \textbf{135} & \textbf{264} & \textbf{-} & \textbf{43} & \textbf{249} \\ \hline
          N detections & $>$ $3\times10^5$ & 310 & 1137 & $>$ $1\times10^5$ & 110 & 612 \\
          Cut in mag & 1156 & 115 & 228 & 792 & 42 & 236 \\
          Cut in C & 891 & 115 & 228 & 621 & 42 & 236 \\
          Cut Gaia & 784 & 115 & 228 & 494 & 42 & 236 \\
          Cut in TF & 273 & 115 & 228 & 258 & 42 & 236 \\ \hline
         \textbf{Final sample (r $<21.3$ mag)} & \textbf{273} & \textbf{115 (85$\%$)} & \textbf{228 (86\%)} & \textbf{258} & \textbf{42 (98\%)} & \textbf{236 (95\%)}\\ \hline
     \end{tabular}
     \label{tab:criteria_GCs}
 \end{table*}

\begin{figure*}
    \centering
        \includegraphics[width=0.85\textwidth]{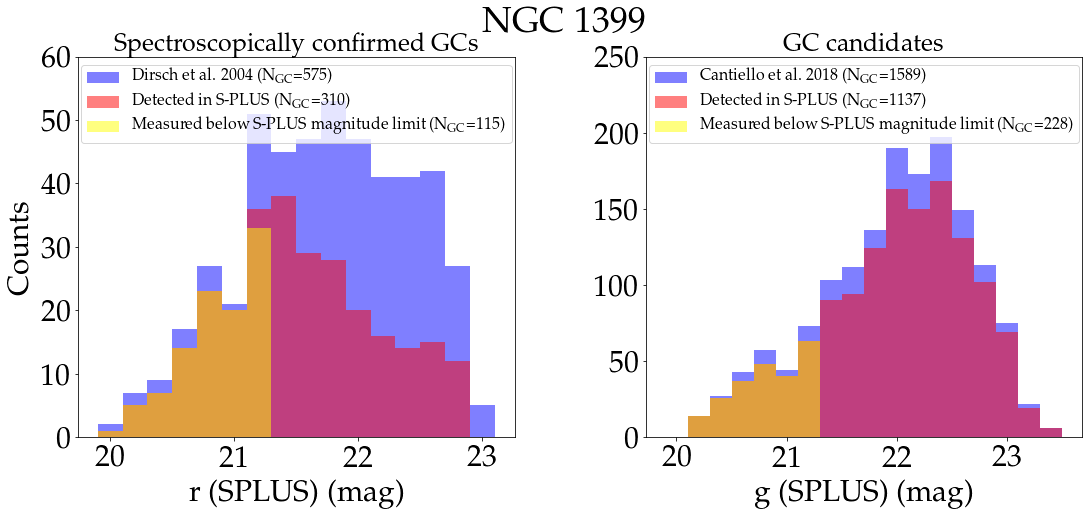}
    \hspace*{-1.0cm}
    \includegraphics[width=0.92\textwidth,trim=0 0 0 2.23cm, clip]{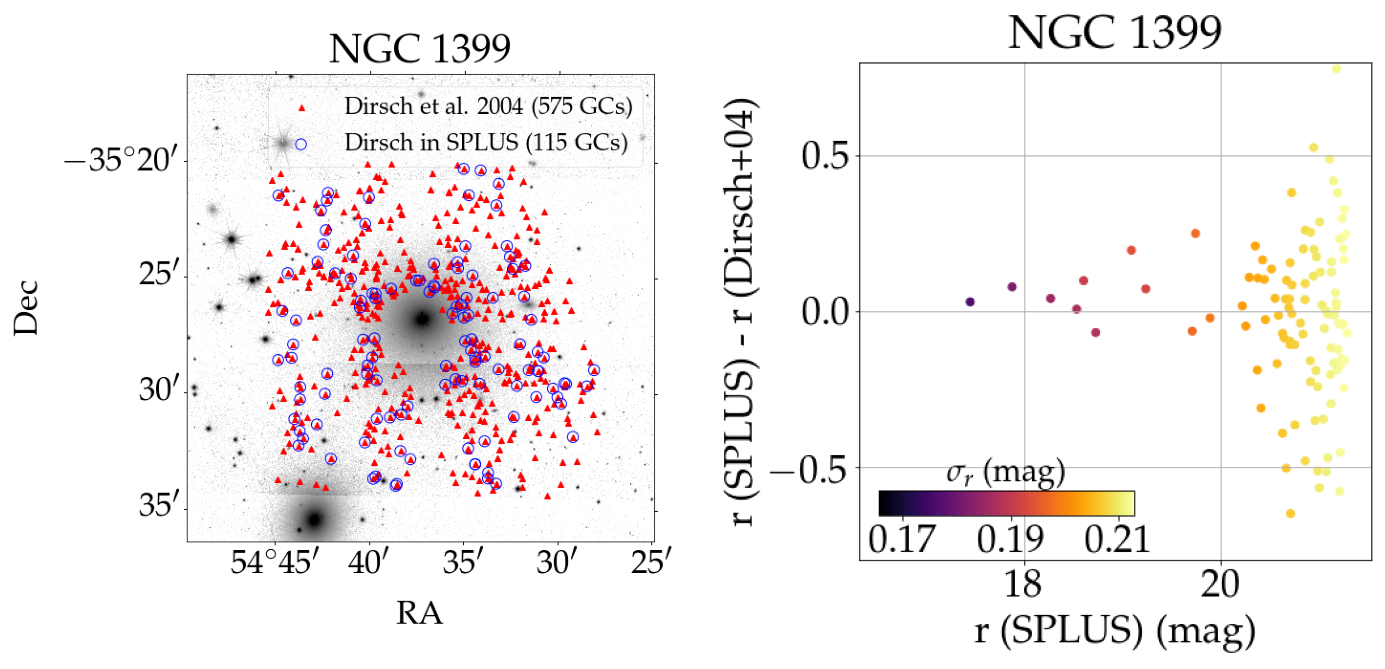}
    \includegraphics[width=0.86\textwidth,trim=0 0 0 1.8cm, clip]{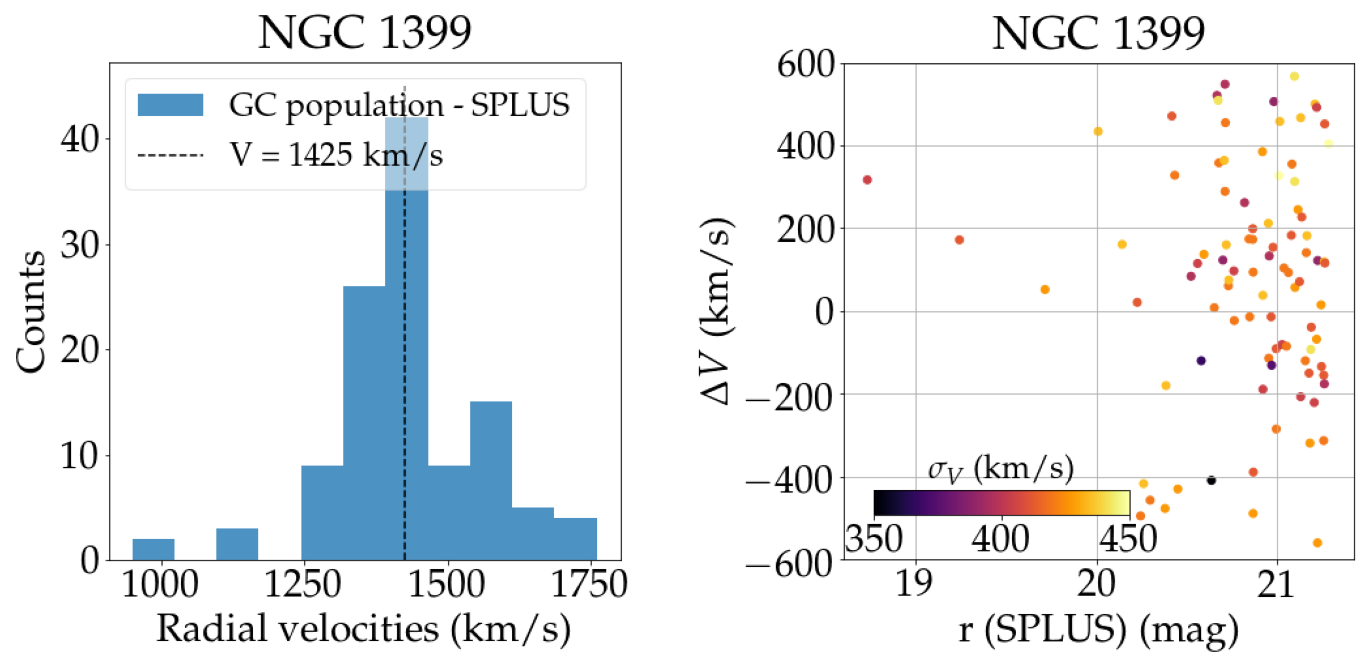}
    \caption{Main properties of the GC candidates around NGC 1399 compared to \protect\cite{Dirsch04} and \protect\cite{Cantiello18}. First row, left: Distribution of all spectroscopically confirmed GCs in purple, of those detected in S-PLUS in pink and of those detected that are below the magnitude limit of S-PLUS ($r<21.3$ mag) in yellow. Right: Same comparison as in the left, but now with the photometric catalog of GC candidates of \protect\cite{Cantiello18}.
   Second row, left: S-PLUS image of NGC 1399. Red triangles show all spectroscopically confirmed GCs, while blue circles show the portion of these confirmed GCs measured with S-PLUS. Right: Photometric comparison between S-PLUS and literature, points are coloured with their photometric errors in S-PLUS. Third row, left: Distribution of GC population radial velocity compared to systemic velocity of NGC 1399 (vertical dashed line). Right: Absolute difference between the radial velocities of the confirmed GCs derived using S-PLUS and using \protect\cite{Dirsch04}. Points are coloured with their radial velocity errors as derived with the template fitting technique.}
    \label{fig:NGC1399}
\end{figure*}

NGC 1399 is the central elliptical galaxy in the Fornax cluster and has a large GC population \citep{Cantiello18}. Thus, it has a GC system extending out to very large radii and with a very diverse population due to its history of sequential mergers and accretions.
We assume that NGC 1399 is at a distance of 19 Mpc \citep{Richtler04} and has an effective radius of 49 arcsec \citep{Iodice16}. To compare our results, we used the spectroscopic sample of \cite{Dirsch04}, which contains 575 confirmed GCs in a composed FoV of $\sim 14 \times 14$ arcmin, and the photometric study of \cite{Cantiello18}, containing 1589 GC candidates, identified using the VEGAS telescope and Fornax Deep Survey data \footnote{However, any comparison with photometric GC candidate lists in literature must be done carefully, as they are comprised of GC candidates only, and false detections might be included in the reference sample as well. \label{foot:note1}}  .

In Table \ref{tab:criteria_GCs}, we show the number of confirmed GCs and GC candidates lost at every step of the analysis with respect to the spectroscopic reference sample and the photometric one, respectively. It is clear that the only cut where we lose GCs (and candidates) is the cut in magnitude. The other cuts (concentration, Gaia and TF) help creating a final list of GC candidates by constraining the initial list of possibilities. 

Additionally, in the first row of Fig. \ref{fig:NGC1399}, we show the number of spectroscopically confirmed GCs and literature GC candidates in purple, compared to those measured in S-PLUS in pink and those that meet the magnitude limit criteria in yellow. As it can be seen, both in Table \ref{tab:criteria_GCs} and on the yellow histogram of Fig. \ref{fig:NGC1399}, we can recover more than 85\% of the confirmed GCs and GC candidates down to $r<21.3$ mag at the distance of 19 Mpc, revealing a high completeness down to the S-PLUS magnitude limit.

When using a new method to detect GC candidates, it is important though to analyse not only the completeness, but also the false positive rate. However, it is hard to discuss false positive rate in a GC candidate list, since without spectroscopy, we cannot know for sure how many detections were incorrectly classified as GCs in our sample. It is even harder to analyse the effect of the newly proposed cuts in the analysis (Gaia, concentration and TF) since it is not possible to perform a similar analysis in the reference catalogues. 
Nonetheless, in order to have an upper-bound estimate of the false positive rate, we did an analysis of a blank field, looking for GC candidates and applying the same selection criteria summarized in Table \ref{tab:selectGCs}. This test was applied to an area of 5 arcmin squared, far away from the studied galaxies (or any known galaxy clusters). After applying each of the criteria, we end up with 3 GC candidates brighter than $r=21.3$ mag in this field. These candidates are likely to be contaminants, i.e., foreground stars or background galaxies. The contaminants have a surface density of 0.12 per arcmin squared. Extrapolating this to the analysed FoV of NGC 1399 (approximately 8 times bigger), we estimate that among our detections, there should be around 24 false positives, representing $\sim$9\% of our total number of GC candidates.

Putting all of this into numbers, as shown in Table \ref{tab:criteria_GCs}, when applying all of the criteria simultaneously, we find 273 GC candidates around NGC 1399. Of those, 115 are spectroscopically confirmed GCs present in the list of \cite{Dirsch04}, and 228 out of the 273 are present in the list of GC candidates compiled from literature in \cite{Cantiello18} \footnote{all of the 115 spectrocopically confirmed GCs are part of the 228 GC candidates}.

Finally, we encounter 45 new GC candidates around NGC 1399. Only through follow up spectroscopy is it possible to confirm these sources as GCs with no doubt, and ultimately assess fully our false positive rate.

After identifying our final GC candidate list, we focus only on the spectroscopically confirmed subsample, which is the sample that we use to measure the radial velocities and stellar populations and that we compare to the literature. In the second row of Fig. \ref{fig:NGC1399}, we show the 2D spatial distribution of the 115 confirmed GCs around NGC 1399 and a comparison of the photometry obtained with S-PLUS and \cite{Dirsch04}. Additionally, for these 115 sources, we were able to recover radial velocities, as shown in the third row of Fig. \ref{fig:NGC1399}, where we compare the distribution of RVs to the systemic velocity of NGC 1399 \citep[1425 km/s, ][]{Richtler04}, and to each individual GC, by measuring the absoulte difference between the radial velocity recovered by us and by \cite{Dirsch04} ($\Delta V$). From Fig. \ref{fig:NGC1399}, we can see that our results are consistent with the literature both for the photometry and systemic velocity of the galaxy. 
Scatter in the radial velocities at $r \sim 21.3$ mag are of the order of $\sim 400$ km/s and magnitude scatter of the order of 0.2 mag at $r \sim 21.3$ mag. 
To compare our results to those in literature, we consider the quadratic sum of the errors derived in this work and the ones tabulated in \cite{Dirsch04}. We find a resulting scatter of 0.5 mag at r $\sim$ 21.3 mag, indicating that the errors obtained from aperture photometry alone may be underestimated by a factor of $\sim$2. However, this analysis of the quadratic sum of the errors is only possible for the r band (the only one in common with \citealt{Dirsch04}) and not for the other bands, especially the narrow ones, since they are unique to the S- PLUS survey. For this reason we use the uncertainty obtained with aperture photometry to perform the SED fitting, in order to use an homogeneously determined error in all bands.

We note that we get a difference in radial velocities of $400-600$ km/s with respect to the reference catalogue and an error on the measured velocities, through SED fitting,  of $250-450$ km/s, this indicates that using an underestimated error in the SED fitting may result in underestimated errors of a few hundred km/s in the radial velocity.

Nevertheless, even if errors in velocity and magnitude might be high on an object-by-object basis, the great advantage of using S-PLUS data, is the possibility of characterizing GC populations as a whole out to very large radii and derive ensemble properties, such as the systemic velocity of galaxies and GCs sub-populations (e.g., blue/red GCs).

\subsection{NGC 3115}

NGC 3115 is the closest lenticular galaxy to the Milky Way (MW) and has a very well studied GC system, with well-established colour bimodalities \citep[e.g.,][]{Brodie14}. 
This galaxy is the only one within the SAGES Legacy Unifying Globulars and GalaxieS survey \citep[SLUGGS,][]{Brodie14} that was observed with S-PLUS and, thus, a great nearby test to our method. 
We adopt the distance of D $= 9.4$ Mpc \citep{Brodie14} and effective radius of 26 arcsec \citep{Cortesi13}. To compare our results, we use the sample of 122 spectroscopically confirmed GCs from \cite{Pota13} (P13), as well as the list of 781 GC candidates of \cite{Jennings14} (J14) (see footnote \textsuperscript{\ref{foot:note1}}).

In Table \ref{tab:criteria_GCs}, we show the amount of objects that were identified in the entire FoV of NGC 3115; we compare this with the number of spectroscopically-confirmed GCs that we lose after applying each one of the criteria listed in Table \ref{tab:selectGCs}, as well as the amount of photometrically-identified GC candidates found by J14 that we lose with each cut.

\begin{figure}
    \centering
    \includegraphics[width=\columnwidth]{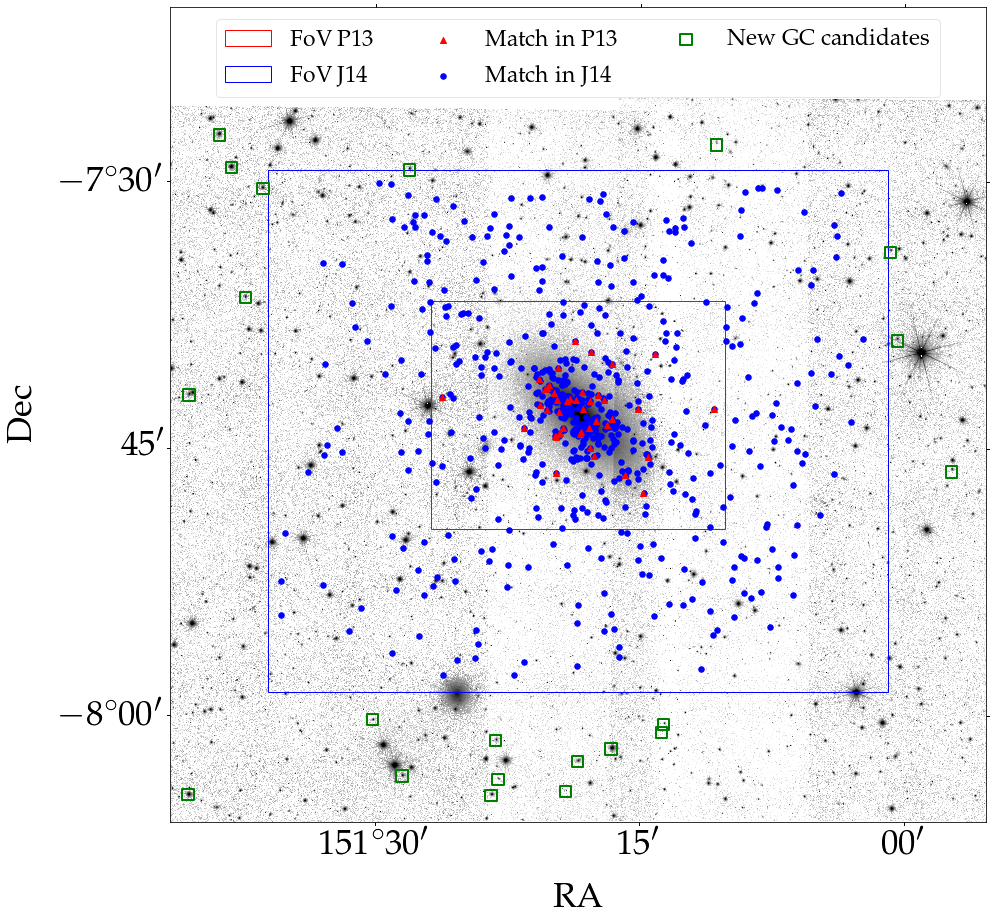}
    \caption{Distribution of GC candidates around NGC 3115. In red, the GCs found in this work that are present in the list of confirmed GCs from P13, and the red rectangle shows the field-of-view of the study of P13. As blue circles, we show the GC candidates found in this work and present in the GC candidate list of J14, while the blue rectangle shows the FoV of the work of J14. The green squares represent the new 22 GC candidates found in this work.}
    \label{fig:distribution_GCs_NGC3115}
\end{figure}

As shown in Table \ref{tab:criteria_GCs}, when applying all of the criteria simultaneously, we find 258 GC candidates around NGC 3115. Out of those, 42 are spectrocopically confirmed GCs (P13) and 236 are already proposed as GC candidates by J14, resulting in 22 new GC candidates found in this work. In Fig. \ref{fig:distribution_GCs_NGC3115}, we show the distribution of the GC candidates found in this work around NGC 3115, comparing the fields of view of our work with that of P13 and J14. In red, we show the GCs that match the spectroscopic sample of P13, in blue, those that match the GC candidate list of J14 and in green the 22 new GC candidates found in this work, where we can clearly see that the new GCs fall outside of the FoV of P13 and J14. This highlights the great advantage of S-PLUS in providing coverage out to larger radii.
Implementing the same analysis described in Section \ref{sec:ngc1399} to estimate the false positive rate (i.e., extrapolating the rate of false detections on a blank field to the field of the studied galaxy), we see that around the FoV of NGC 3115 (approximately 20 arcmin squared), there are 12 GC candidates that might be false detections, which would represent 5\% of our total number of GC candidates. We remind the reader that this is an upper bound estimate of the false positive rate and only through spectroscopy we can assess the truth rate.

\begin{figure*}
    \centering
    \includegraphics[width=0.85\textwidth]{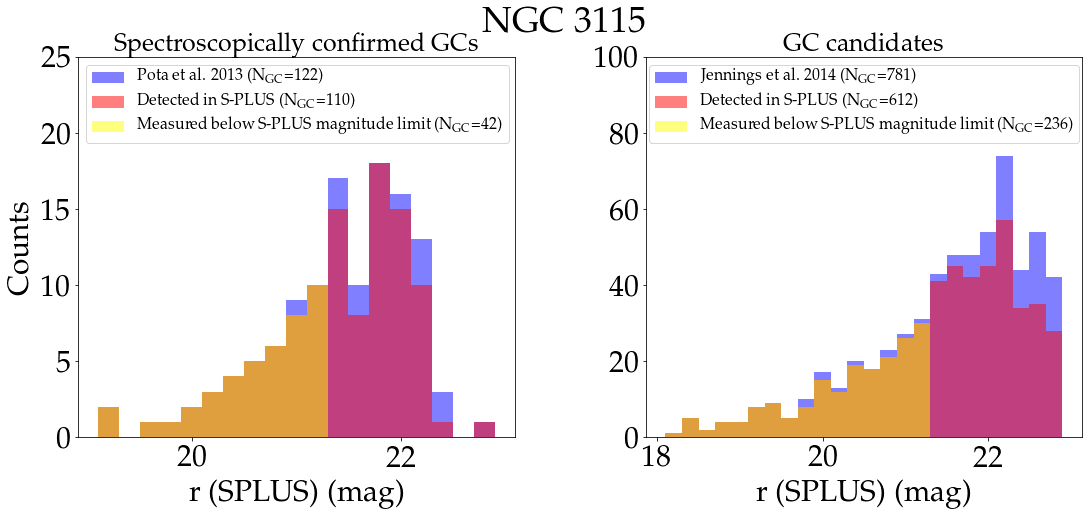}
    \hspace*{-1.0cm}
    \includegraphics[width=0.92\textwidth,trim=0 0 0 1.75cm, clip]{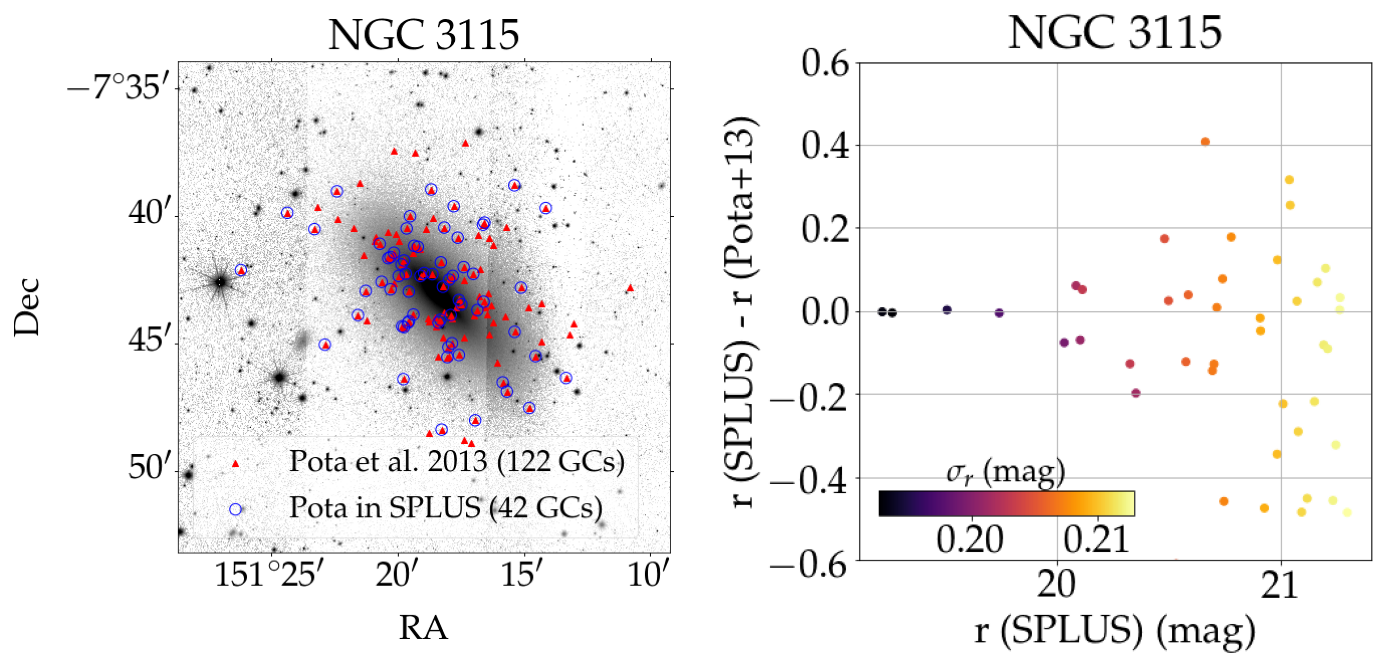}
    \includegraphics[width=0.86\textwidth,trim=0 0 0 2.23cm, clip]{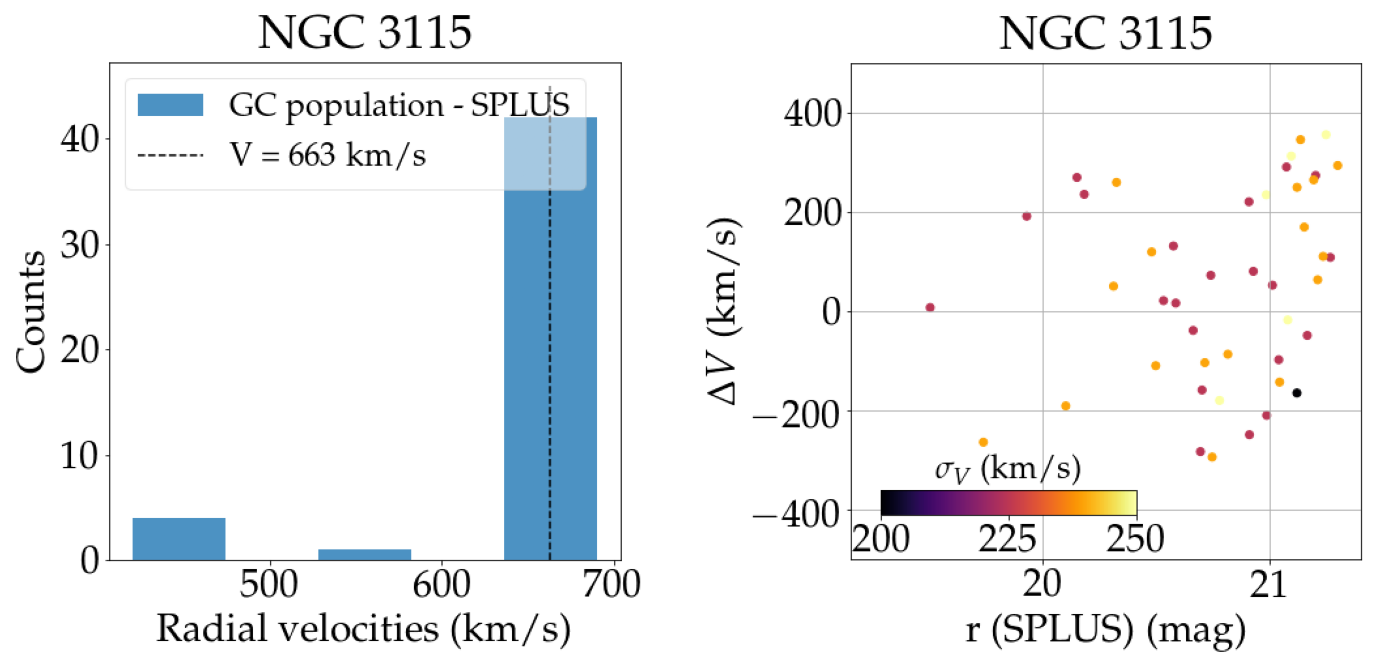}
    \caption{Main properties of the GC candidates around NGC 3115 compared to \protect\cite{Pota13} and \protect\cite{Jennings14}. First row, left: Distribution of all spectroscopically confirmed GCs in purple, of those detected in S-PLUS in pink and of those detected that are below the magnitude limit of S-PLUS ($r<21.3$ mag) in yellow. Right: Same comparison as in the left, but with the photometric catalog of GC candidates of J14.
    Second row, left: S-PLUS image of NGC 3115. Red triangles show all spectroscopically confirmed GCs, while blue circles show the portion of these confirmed GCs measured with S-PLUS. Right: Photometric comparison between S-PLUS and literature, points are coloured with their photometric errors in S-PLUS. Third row, left: Distribution of GC population radial velocities compared to the systemic velocity of NGC 3115 (vertical dashed line). Right: Absolute difference between the radial velocities derived with S-PLUS and those in P13. Points are coloured with their radial velocity errors as derived with the template fitting technique.}
    \label{fig:NGC3115}
\end{figure*}

In the first row of Fig. \ref{fig:NGC3115}, we show the number of spectroscopically confirmed GCs and literature GC candidates in purple, compared to those measured in S-PLUS in pink and those that meet the magnitude limit criteria in yellow. As it can be seen, both the first row of Fig. \ref{fig:NGC3115} and Table \ref{tab:criteria_GCs} show that we are able to detect more than 95\% of the GC candidates from J14 and 98\% of the GCs from P13, showing that the identification and characterisation of GCs with S-PLUS is near to complete down to $r \sim 21.3$ mag at the distance of 10 Mpc.
We lose 13 GC candidates from J14 out of the 249, corresponding to less than 5\% of the detections, and 1 confirmed GC from P13 out of the 43, since these GCs are located in the bright central part of the galaxy, as it can be seen in Fig. \ref{fig:NGC3115}.

When compared to NGC 1399, we see that our completeness around NGC 3115 is slightly higher, given that at the distance of NGC 1399, many objects surpass the limitations of S-PLUS. 

To study the stellar populations and radial velocities, we now turn ourselves only to the sample of 42 spectroscopically confirmed GCs around NGC 3115. In the second row of Fig. \ref{fig:NGC3115}, we show the 2D spatial distribution of confirmed GCs around NGC 3115, and compare the photometry obtained with S-PLUS and with P13. In the third row on the left, we compare the distribution of radial velocities of the GCs to the systemic velocity of NGC 3115 \citep[663 km/s, ][]{Pota13}. And in the right, we compare the RVs of each GC with the ones obtained by P13, by looking at the absolute difference between the values derived in both works ($\Delta V$). Typical scatter in the radial velocity estimates at $r \sim 21.3$ mag are of the order of 200 km/s and scatter in the photometric errors also at $r \sim 21.3$ mag are about 0.2 mag.

\begin{figure*}
    \centering
    \includegraphics[width=0.96\columnwidth]{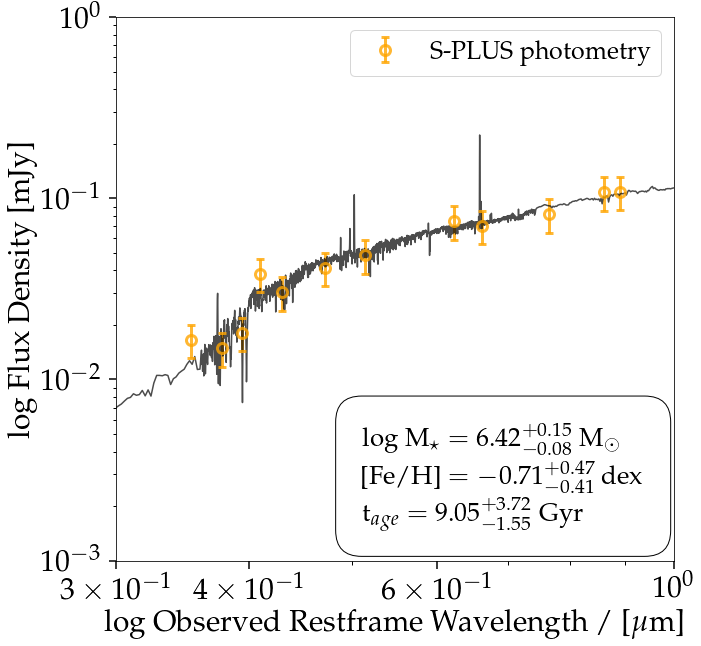}
    \includegraphics[width=1.1\columnwidth,trim = 0 0.5cm 12cm 1cm, clip]{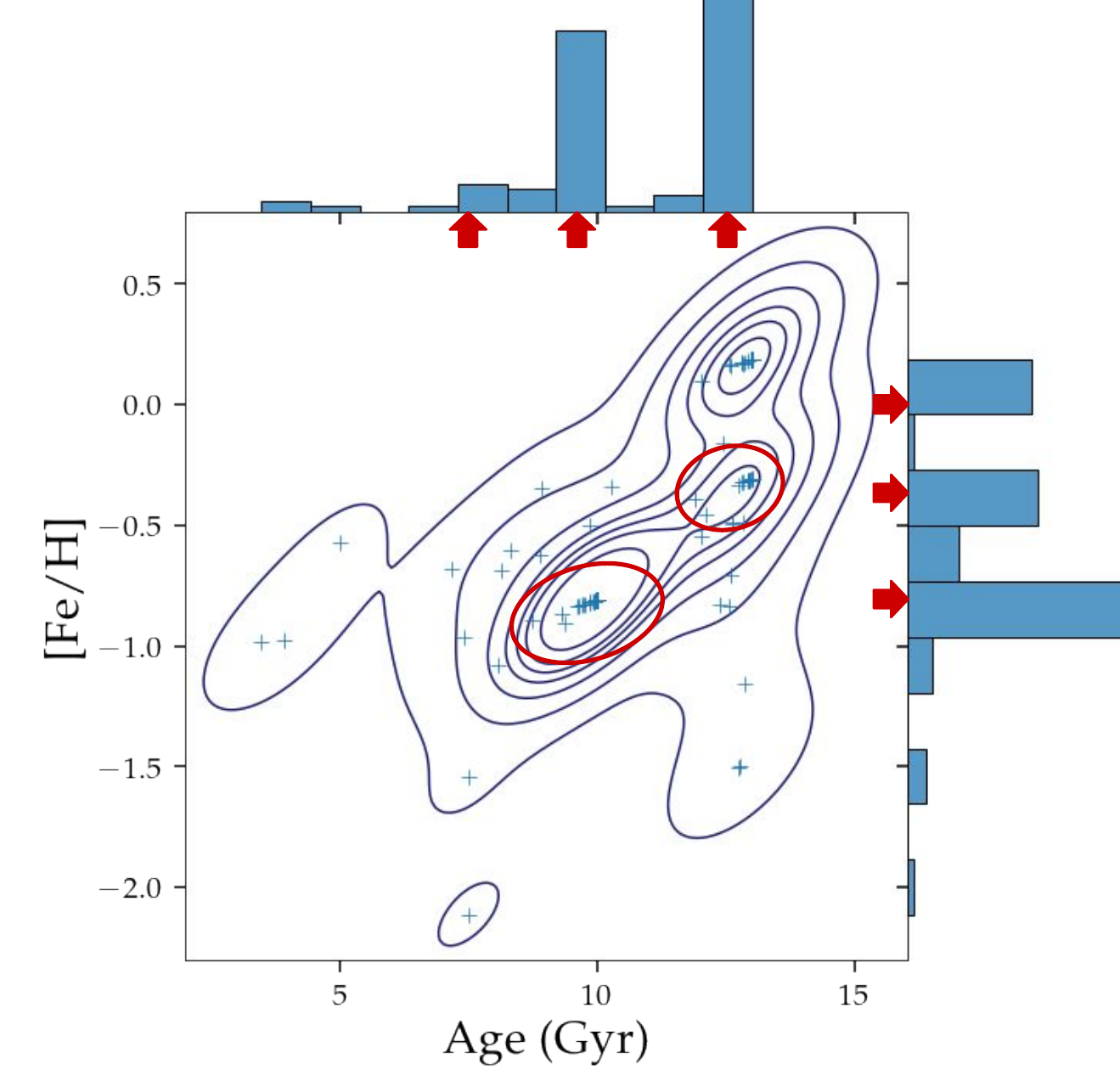}
    \caption{SED fitting results of confirmed GCs around NGC 3115. Left: Orange points show the S-PLUS magnitudes, and black curve the best-fit SED. Inset box summarises the best-fit parameters (stellar mass, metallicity and age). Right: Distribution of age and [Fe/H] retrieved with SED fitting for the GC population around NGC 3115. Blue contours show the density of objects and marginal histograms show the distribution of each parameter. The red arrows indicate the approximate centre of the peaks of the distribution of GCs ages and metallicities in \protect\cite{Usher}, whereas the red ellipses show the centre of the contours of the results of \protect\cite{Usher}, showing that the derived stellar populations are consistent within errors.}
    \label{fig:SED3115}
\end{figure*}

To analyse our measured stellar population properties, we use the study of \cite{Usher}, that recovered ages and metallicities for 116 GCs around NGC 3115 using Keck/DEIMOS spectra.
In the left panel of Fig. \ref{fig:SED3115} we show an example of SED fitting for one of the GCs around NGC 3115, including a summary of the marginalized posteriors of the stellar mass, metallicity and age. In the right panel, we show the age (typical scatter of 2 Gyr) and metallicity (typical uncertainty $<$0.5 dex) distribution of the entire GC population around NGC 3115.

The SED fitting results shown in Fig. \ref{fig:SED3115} are consistent with the ages and metallicities derived by \cite{Usher} for the GC system around NGC 3115, as shown by the arrows and ellipses in Fig. \ref{fig:SED3115} (see Fig.4 of \citealt{Usher} for a more detailed analysis), assuming a scatter of 0.5 dex in metallicity and 2 Gyr in age. In particular, we see that GCs with higher metallicities ([Fe/H]$> -0.8$ dex) compose the older GC populations ($\overline{\rm age} = 13$ Gyr), while GCs with [Fe/H]$<-0.8$ dex have younger mean ages ($\overline{\rm age} = 9$ Gyr). When comparing the distribution of the ages and metallicities, we can see that our method recovers three peaks in metallicity, similar to those of \cite{Usher}. Regarding ages, differently from \cite{Usher}, we recover two peaks, instead of three, although the recovered two in $\sim 9$ and $\sim 13$ Gyr are also present in \cite{Usher}. Although we do not recover the peak of the younger GCs, as in \cite{Usher} ($\sim 7$ Gyr), we do find three GCs in this range, as it can be seen in Fig. \ref{fig:SED3115}, but these are too few to configure a new peak in the distribution. It is expected that we would not recover the exact same distribution though, since we are using only the 42 GCs, while \cite{Usher} uses a sample of 116 GCs. This shows, nonetheless, that using only photometric data from S-PLUS, we are able to recover stellar populations properties of the GCs similar to those obtained using spectroscopic data \citep{Usher}.

Finally, we note that although errors in the age and metallicity are large for individual measures; statistically, the method is very powerful in identifying GC populations around galaxies and deriving their ensemble properties, even for sub-populations (e.g., blue/red GCs).

\section{Conclusions}
In this work, we probed the ability of the S-PLUS survey to find and characterize extragalactic GCs. We developed a new method to carefully select GC candidates among all detections, excluding stars and compact galaxy contaminants. For this, we use a combination of photometric and concentration selection criteria, template fitting techniques and Gaia's proper motions. After applying our selection cuts, we effectively verify our ability to identify almost the entirety of the GCs previously confirmed in literature down to $r\sim21.3$ mag, which corresponds to the SPLUS survey 3$\sigma$ depth. This corresponds to 42 objects, down to M$_V$ $<$ -8.8 mag around NGC 3115 and 115 objects, down to M$_V$ $<$ -10.1 mag around NGC 1399. With this survey depth, we note that GCs in galaxies in the local Universe out to 5.7 Mpc (e.g., M81, NGC 253, CenA, etc.) would be selected down to the turnover magnitude of the globular cluster luminosity function (GCLF, M$_V = -7.5$ mag). 

We show that we can measure physical properties of the GCs using SED fitting, such as metallicity, mass and age. In particular, we recover the age and metallicity distributions, similar to what was previously measured spectroscopically, in the case of NGC 3115. 
Although it is not the goal of this paper, we note that, using this technique, we find a number of new GC candidates in both systems, with $r < 21.3$ mag, whose spectroscopy has not been obtained yet. We plan to obtain spectroscopy of those to further validate the technique, specially for GCs in the very outskirts of the galaxy.

We conclude that the filter system used by S-PLUS and J-PLUS, a combination of narrow and broad band filters, combined with a focused pipeline to select and characterise GCs, allows for the recovery of ensemble properties of the GC populations over a wide FoV, overcoming the current incomplete area coverage limitation of spectroscopic samples of GCs. Expanding this study to the entire Southern sky (and Northern, with J-PLUS), including several thousand galaxies in different environments and distances, as well as unbound GCs lying in the intracluster medium, will provide an unprecedented overview of GC populations down to r$\sim21.3$ mag.

\section*{Acknowledgements}
We deeply thank the anonymous referee for the insightful suggestions and comments, which resulted in great additions to the paper. We thank Roderik Overzier for the help with the template fitting technique.
MLB and CMdO acknowledge the financial support of the Sao Paulo Research Foundation (FAPESP) under grant 2019/23388-0. CEB acknowledges FAPESP, grant 2016/12331-0.
PC acknowledges support from Conselho Nacional de Desenvolvimento Cient\'ifico e Tecnol\'ogico (CNPq) under grant 310041/2018-0.
D.B.-S. also acknowledges Funda\c{c}\~ao de Amparo \`{a} Pesquisa do Estado de S\~ao Paulo (FAPESP) process number 2017/00204-6 for the financial support.
ACS acknowledge funding from the brazilian agencies \textit {Conselho Nacional de Desenvolvimento Cient\'ifico e Tecnol\'ogico} (CNPq) and the  \textit{Fundação de Amparo à Pesquisa do Estado do RS} (FAPERGS) through grants CNPq-403580/2016-1, CNPq-11153/2018-6, PqG/FAPERGS-17/2551-0001, FAPERGS/CAPES 19/2551-0000696-9, L'Or\'eal UNESCO ABC \emph{Para Mulheres na Ci\^encia} and the Chinese Academy of Sciences (CAS) President's International Fellowship Initiative (PIFI) through grant E085201009.
AAC acknowledges support from the State Agency for Research of the Spanish MCIU through the ``Center of Excellence Severo Ochoa'' award to the Instituto de Astrofísica de Andalucía (SEV-2017-0709). This work was funded with grants from Consejo Nacional de Investigaciones Científicas y Técnicas and Universidad Nacional de La Plata (Argentina).
The S-PLUS project, including the T80-South robotic telescope and the S-PLUS scientific survey, was founded as a partnership between the Fundação de Amparo à Pesquisa do Estado de São Paulo (FAPESP), the Observatório Nacional (ON), the Federal University of Sergipe (UFS), and the Federal University of Santa Catarina (UFSC), with important financial and practical contributions from other collaborating institutes in Brazil, Chile (Universidad de La Serena), and Spain (Centro de Estudios de Física del Cosmos de Aragón, CEFCA). We further acknowledge financial support from the São Paulo Research Foundation (FAPESP), the Brazilian National Research Council (CNPq), the Coordination for the Improvement of Higher Education Personnel (CAPES), the Carlos Chagas Filho Rio de Janeiro State Research Foundation (FAPERJ), and the Brazilian Innovation Agency (FINEP).
The authors are grateful for the contributions from CTIO staff in helping in the construction, commissioning and maintenance of the T80-South telescope and camera. We are also indebted to Rene Laporte and INPE, as well as Keith Taylor, for their important contributions to the project. We also thank CEFCA staff for their help with T80-South, specifically we thank Antonio Marín-Franch for his invaluable contributions in the early phases of the project, David Cristóbal-Hornillos and his team for their help with the installation of the data reduction package jype version 0.9.9, César Íñiguez for
providing 2D measurements of the filter transmissions, and all other staff members for their support.

\section*{Data Availability}
The S-PLUS DR3 data are available via the S-PLUS archive (\href{https://splus.cloud/}{splus.cloud}) for the general public from December 1st, 2021.



\bibliographystyle{mnras}
\bibliography{bibli} 





\bsp	
\label{lastpage}
\end{document}